\documentclass[10pt,pre,aps,twocolumn,superscriptaddress,floatfix,notitlepage,longbibliography]{revtex4-1}
\usepackage[latin1]{inputenc}
\usepackage{mathrsfs}
\usepackage{amsmath,color}
\usepackage{amsfonts,natbib}
\usepackage{amssymb}
\usepackage{bm}
\usepackage[pdftex]{graphicx}

\usepackage{graphicx}

\begin{document}
	\title{Efficiently Fuelling a Quantum Engine with Incompatible Measurements} 
	\author{Sreenath K. Manikandan}
	\email{sreenath.k.manikandan@su.se}
	\affiliation{Department of Physics and Astronomy, University of Rochester, Rochester, NY 14627, USA}
	\affiliation{Center for Coherence and Quantum Optics, University of Rochester, Rochester, NY 14627, USA}
	\affiliation{Nordita,
KTH Royal Institute of Technology and Stockholm University,
Hannes Alfv\'{e}ns v\"{a}g 12, SE-106 91 Stockholm, Sweden}
	\author{Cyril Elouard}
	\affiliation{Department of Physics and Astronomy, University of Rochester, Rochester, NY 14627, USA}
	\affiliation{QUANTIC lab, INRIA Paris, 2 Rue Simone Iff, 75012 Paris, France}
	\author{Kater W. Murch}
	\affiliation{Department of Physics, Washington University, St. Louis, Missouri 63130}
	\author{Alexia Auff\`{e}ves}
	\affiliation{Universit\'{e} Grenoble Alpes, CNRS, Grenoble INP, Institut N\'{e}el, 38000 Grenoble, France}
	\author{Andrew N. Jordan}	\affiliation{Institute for Quantum Studies, Chapman University, Orange, CA, 92866, USA}
	\affiliation{Center for Coherence and Quantum Optics, University of Rochester, Rochester, NY 14627, USA}    
\affiliation{Department of Physics and Astronomy, University of Rochester, Rochester, NY 14627, USA}
\date{\today}
	\begin{abstract}
We  propose a quantum harmonic oscillator measurement engine fueled by simultaneous quantum measurements of the non-commuting position and momentum quadratures of the quantum oscillator. The engine extracts work by moving the harmonic trap suddenly, conditioned on the measurement outcomes. We present two protocols for work extraction, respectively based on single-shot and time-continuous quantum measurements.  In the single-shot limit, the oscillator is measured in a coherent state basis; the measurement adds an average of one quantum of energy to the oscillator, which is then extracted in the feedback step.  In the time-continuous limit, continuous weak quantum measurements of both position and momentum of the quantum oscillator result in a coherent state, whose coordinates diffuse in time. We relate the extractable work to the noise added by quadrature measurements, and present exact results for the work distribution at arbitrary finite time. Both protocols can achieve unit work conversion efficiency in principle. 
	\end{abstract}
	\maketitle
\section{Introduction}Quantum thermodynamics is concerned with how the exchange of heat and work can be understood and applied when quantum effects such as entanglement and coherences are present~\cite{vinjanampathy_quantum_2016,goold_role_2016,binder_thermodynamics_2019,jordan_quantum_2020}. The emerging field of ``quantum energetics" is applicable to stochastic energy exchanges when there are no thermal baths. As an example, quantum measurement powered engines have been proposed with qubit systems, as well as continuous variable systems~\cite{elouard_efficient_2018,jacobs_second_2009,ding_measurement-driven_2018,solfanelli_maximal_2019,elouard_interaction-free_2020,das_measurement_2019,debarba_work_2019,yi_single-temperature_2017,buffoni_quantum_2019,elouard_extracting_2017,seah_maxwells_2020,bresque_two-qubit_2021}. This line of research is greatly stimulated by successful demonstrations of quantum measurements and control in a variety of quantum platforms including but not limited to superconducting circuits~\cite{vijay_stabilizing_2012,campagne-ibarcq_persistent_2013,riste_deterministic_2013}, cavities~\cite{sayrin_real-time_2011,zhou_field_2012,steck_quantum_2004}, trapped ions~\cite{leibfried_quantum_2003,schindler_experimental_2011,bushev_feedback_2006}, trapped nano-particles~\cite{gieseler_subkelvin_2012}, single electron systems~\cite{durso_feedback_2003}, and mechanical resonators~\cite{rossi_measurement-based_2018,sudhir_appearance_2017}. 

By making appropriate combination of measurements and feedback operations~\cite{masuyama_information--work_2018,leff_maxwells_2014,cottet_observing_2017,naghiloo_information_2018,camati_experimental_2016,koski_experimental_2014,koski_-chip_2015,klatzow_experimental_2019}, a quantum engine can be used to accelerate an electron to charge a capacitor, or to lift a tiny mass~\cite{elouard_efficient_2018}. A quantum refrigerator based on measurements~\cite{buffoni_quantum_2019}, measurement driven single temperature engines that require no feedback~\cite{yi_single-temperature_2017,ding_measurement-driven_2018}, interaction-free measurement engines~\cite{elouard_interaction-free_2020}, and quantum measurement engines driven by quantum entanglement~\cite{bresque_two-qubit_2021} have also been conceptualized, extending the scope of measurement based thermal-equivalent machines. The prime focus  in these models has been on the measurement of a single observable, either the spin along a chosen axis in the case of finite dimensional systems~\cite{solfanelli_maximal_2019,elouard_extracting_2017,elouard_role_2017,bresque_two-qubit_2021}, or a given quadrature with continuous variable systems~\cite{elouard_efficient_2018,ding_measurement-driven_2018,opatrny_work_2021}. 
 \begin{figure}
\includegraphics[width=\linewidth]{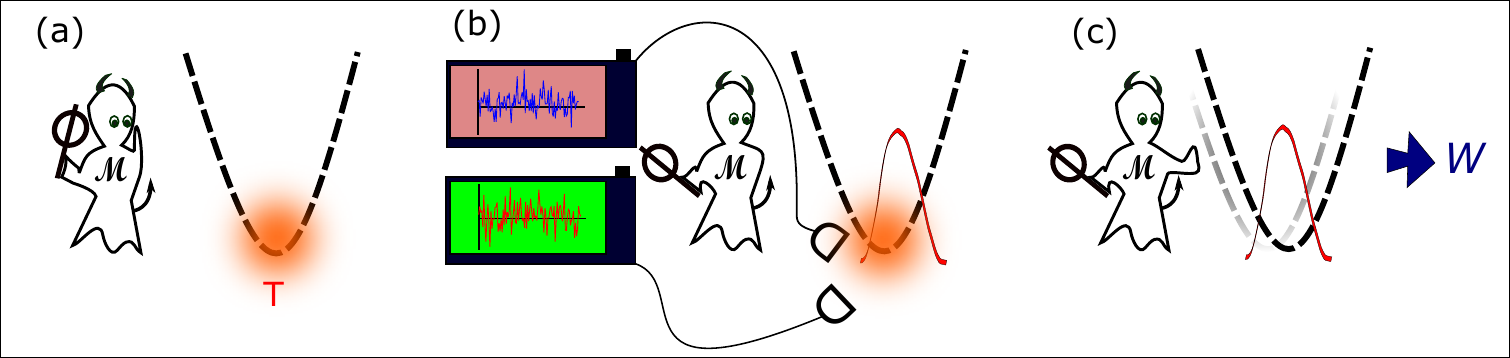}\caption{Cyclic operation of the quantum oscillator measurement engine fueled by incompatible measurements. (a) The quantum oscillator is initially thermalized to a heat bath at temperature T. (b) A demon weakly measures both the position and momentum of the quantum oscillator. The measurement results in a coherent state. (c) Work is extracted by displacing the trap conditioned on the measurement outcomes.
\label{fig1}}
\end{figure}
 
 In this article, we propose a quantum engine fueled by simultaneous weak measurements of two non-commuting observables: the position and momentum of a quantum oscillator. Work is extracted by moving the bottom of the harmonic trap suddenly, conditioned on the measurement outcomes. We show that incompatible measurements have rewarding energetic consequences when compared to similar protocols for a quantum engine fueled by measurement of a single quadrature~\cite{opatrny_work_2021,elouard_efficient_2018}, and when compared to the classical limit which describes a heat engine fueled by measuring the position of a Brownian particle in a harmonic trap~\cite{ashida_general_2014,paneru_lossless_2018}. With simultaneous quantum measurements, the measurement strengths for each of the incompatible quadratures  can be tuned such that the measurement results in a displaced ground state (a coherent state~\cite{sudarshan_equivalence_1963,glauber_coherent_1963}) of the quantum oscillator~\cite{arthurs_simultaneous_1965,karmakar_stochastic_2021}. This allows one to perfectly reset the engine's cycle by work extraction, in the same way as conceived in the original version of the Szilard engine, where the demon resets a gas of particles at no cost: it uses information on the positions to compress the gas on the left of a vacuum chamber, and then lets it expand back to equilibrium~\cite{szilard_uber_1929}.  The engine is efficient because the work extraction step precisely brings it back to the equilibrium state.
 When simultaneous weak measurements of position and momentum are performed on a thermal state, it is also well known that they add an extra quantum of noise in to the oscillator~\cite{arthurs_simultaneous_1965,caves_quantum_2012}. This added noise from measurement allows us to extract work even at zero temperature, beating the previously known classical bounds for a Brownian heat engine (see Appendix A)~\cite{ashida_general_2014,paneru_lossless_2018}.

\section{The setup}A quantum harmonic oscillator is described by the Hamiltonian,
$
    \hat{\mathcal{H}}=\frac{\hat{p}^{2}}{2m}+m\omega^{2}\frac{\hat{x}^{2}}{2},
$
where $\hat{p}$ and $\hat{x}$ obey the commutation relation,~
$
    [\hat{x},\hat{p}] = i\hbar.
$
When the temperature is sufficiently small such that $k_{B}\text{T}\ll\hbar\omega$, thermal fluctuations are negligible and the quantum harmonic oscillator will be in its ground state $|0\rangle$. The premise  of the measurement engine is that a suitable weak quantum measurement will excite the oscillator. Work can then be extracted by a feedback loop that changes the harmonic trap suddenly.  The feedback is most efficient if it resets the quantum engine to its initial quantum state at the end of each cycle, making the engine prepared for the next cycle to begin. For the quantum oscillator, a simultaneous weak quantum measurement of both position and momentum observables is uniquely suited to this task because such a protocol realizes measurements in the coherent state basis~\cite{sudarshan_equivalence_1963,glauber_coherent_1963,arthurs_simultaneous_1965,karmakar_stochastic_2021}.

We now proceed to describing two protocols for optimal work extraction with incompatible quantum measurements of the quantum oscillator: a single-shot measurement protocol where measurements are described by projection onto the coherent state basis~\cite{arthurs_simultaneous_1965}, and the time-continuous limit where continuous weak quantum  measurements of both position and momentum of the quantum oscillator results in a coherent state whose coordinates diffuse in time~\cite{karmakar_stochastic_2021}. 

\subsection{Single-shot quantum measurements}The protocol takes place in three steps (see Fig.~\ref{fig1}):
\begin{itemize}    \item {Step (a): The quantum oscillator thermalizes with the ambient inverse temperature, $\beta =1/k_{B}\text{T}$, yielding a thermal state, $\rho(0)=\frac{e^{-\beta\hat{\mathcal{H}}}}{\mathcal{Z}}$, where $\mathcal{Z}=\text{tr}\{e^{-\beta\hat{\mathcal{H}}}\}$.}
\item {Step (b): The quantum oscillator is weakly measured in the coherent state basis $\{|\alpha\rangle\}$, yielding result $\alpha$.}
\end{itemize}
The Kraus operators describing the measurement are,
 $   K(\alpha)=\frac{1}{\sqrt{\pi}}|\alpha\rangle\langle\alpha|$~\cite{kraus_states_1983,ochoa_simultaneous_2018},
and the normalized probability distribution corresponding to the readouts is,
\begin{eqnarray}
 \mathcal{P}_{\text{Q}}(r,\bar{n}) = \frac{1}{\pi}\langle \alpha|\frac{e^{-\beta\mathcal{H}}}{\mathcal{Z}}|\alpha\rangle=\frac{1}{\pi(1+\bar{n})}e^{-r^{2}/(1+\bar{n})},\label{pq}
\end{eqnarray}
using the parametrization $\alpha = re^{i\theta}$.  We denote the average photon number in the initial thermal state by $\bar{n} = (e^{\frac{\hbar\omega}{k_{B}\text{T}}}-1)^{-1}$, the Bose-Einstein occupation. The probability density $\mathcal{P}_{\text{Q}}(r,\bar{n})$ is also known as the Husimi Q distribution of the state $\rho(0)$~\cite{husimi_formal_1940,caves_quantum_2012} which is obtained when simultaneous weak measurements of position and momentum are performed on a thermal state. The average quanta in $\mathcal{P}_{\text{Q}}(r,\bar{n})$ is $\bar{n}+1$, meaning that the measurement process adds one extra quantum to the oscillator quantum state (see Appendix B). 

We may allow the quantum harmonic oscillator to undergo free evolution, $|\alpha\rangle\rightarrow|\alpha e^{-i\omega\tau(\alpha)}\rangle = |r\rangle$ at the end of which the coherent state is located along the positive $x$ axis in the phase space, $(x,p)$. Here $\tau (\alpha) = \theta/\omega$.
This free unitary evolution is essentially a rectifier which channels an arbitrary displacement to a preferred direction, using quantum feedback. The efficiency of this step will require that the time scale of thermalization, $\tau_{_{\text{T}}}\gg 2\pi/\omega$.

\begin{itemize}
\item{Step (c) (work extraction): We suddenly shift the quantum harmonic trap, such that the coherent state $|r\rangle$ is the new quantum ground state. 
In the process, we extract the amount of work, $W = \hbar\omega r^{2}>0$.
 The system thermalizes to the ambient temperature T, completing the cycle.}
\end{itemize}
In a quantum LC circuit implementation, the work extraction in step (c) would correspond to modifying the offset voltage in the capacitor suddenly. Alternatively, one can also modify the branch flux/current in the circuit, if the free evolution aligns the coherent state along the flux/current axis, or a combination of displacements in both voltage and current by an appropriate choice, $\tau(\alpha)$. An alternate Binary-valued feedback protocol for the engine is presented in Appendix C. 

\subsection{Extractable work}The average amount of work extractable from the quantum harmonic oscillator in steps (a)--(c) is,
\begin{eqnarray}
    \langle W\rangle&=&\hbar\omega\int_{0}^{2\pi} d\theta\int_{0}^{\infty} rdr~r^{2}\mathcal{P}_{\text{Q}}(r,\bar{n}) =\hbar\omega(1+\bar{n}).
\end{eqnarray}
 Note that the extra quantum added by the measurement process  (see Appendix B) is also extracted perfectly in the feedback step.  This additional quantum of energy is a purely quantum effect, which exceeds the classical bound on extractable work from the Brownian heat engine at zero temperature (see Appendix A). Addition of extra noise from simultaneous measurement of non-commuting observables is required by quantum mechanics~\cite{arthurs_simultaneous_1965,loudon_squeezed_1987,clerk_introduction_2010,caves_quantum_1982,caves_quantum_2012,bergeal_phase-preserving_2010,haus_quantum_1962}, and our measurement engine exploits the energetic consequence of this added noise by demonstrating that it can be rectified to produce useful work. 
 

\begin{figure*}
\includegraphics[width=\linewidth]{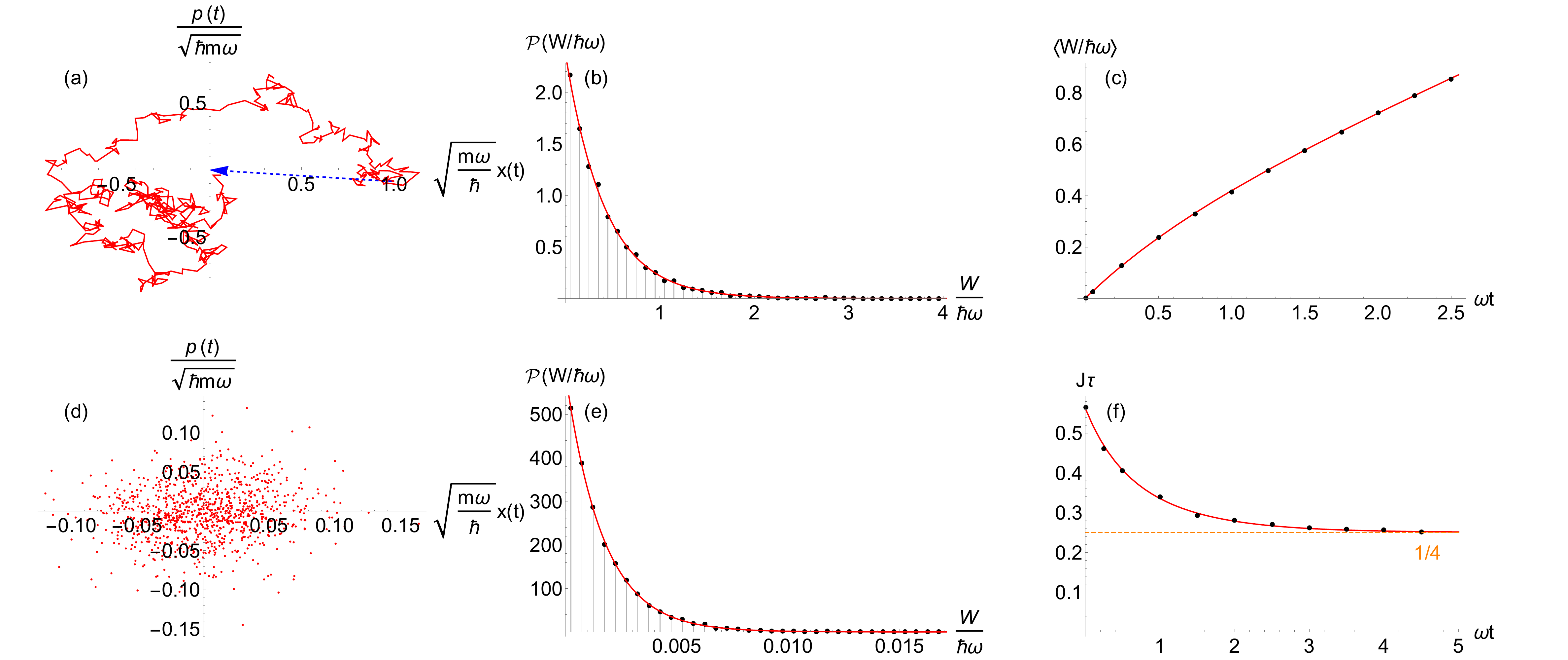}\caption{(a) A single quantum trajectory resulting from continuous quantum measurement of both position and momentum observables, for a duration $\omega t= 2.5$. At time $\omega t= 2.5$ the oscillator is reset (blue arrow). (b) The probability distribution of extractable work if the feedback is applied only once at $\omega t=1$, from simulation of $10^{4}$ trajectories. The red curve is the exact prediction for the probability distribution of extractable work. (c) The average of extractable work when the feedback is performed only once at a given $\omega t$. (d) The average position and momentum of the harmonic oscillator after every single step measurement, for a duration $\omega t=5.$ Here the oscillator is reset to the origin by a feedback applied after each measurement with timestep $\omega dt=0.005$. (e) The probability distribution of extractable work at $\omega t=1$ from simulation of $10^{4}$ trajectories, for continuous feedback applied after every single step measurement. (f) The average power J delivered by the engine per measurement rate $\tau^{-1}$, as a function of duration of the measurement. The engine reaches a steady state with unit efficiency in the large time limit when the quadrature variances of the oscillator (which are identical), $\nu(t)$ tends to $1/2.$ \label{Fig2}}
\end{figure*}
The extracted work is also equal to the sum of the average energy given by the thermal bath ($\mathcal{Q}_{\text{T}}=\hbar\omega\bar{n}$) and the average energy given by the measurement process ($\mathcal{Q}_{M}=\hbar\omega$). Hence energy is conserved and we have $\langle W\rangle=\mathcal{Q}_{\text{T}}+\mathcal{Q}_{M}\equiv \mathcal{Q}$. In addition, energy $\mathcal{Q}_{M}$ is lost by the measuring apparatus such that the latter has to be re-energized to be compatible with the Wigner-Araki-Yanase theorem~\cite{wigner_messung_1995,araki_measurement_1960,yanase_optimal_1961,ahmadi_wignerarakiyanase_2013}. The measurement engine in this case also has unit work conversion efficiency, i.e., $\eta=\langle W\rangle/\mathcal{Q}=1$. Above we have not accounted for the additional cost to erase the memory of the Maxwell's demon, which is discussed in Appendix D~\cite{landauer_irreversibility_1961}.

\section{Continuous weak quantum measurement protocol}We now describe a time-continuous operation of our quantum oscillator measurement engine, where continuous weak quantum measurement of both position and momentum of the quantum oscillator results in a coherent state whose coordinates diffuse in time. Work is extracted by moving the bottom of the harmonic trap; either time-continuously as measurement results are accumulated, or at the end. 

For the engine, we assume that the oscillator is initialized in a Gaussian state, so the first moments together with the covariance matrix elements $V_{ij}=\langle 1/2\{\hat{q}_{i}- q_{i}, \hat{q}_{j}-q_{j}\}\rangle$, completely specify the quantum state~\cite{simon_quantum-noise_1994}. Here $\hat{q}_{1}=\hat{x},~\hat{q}_{2}= \hat{p},~\text{and}~q_{i}=\langle \hat{q}_{i}\rangle,~i=1,2$. Simultaneous weak measurements of both the position and momentum observables of the quantum harmonic oscillator given the continuous readouts $r_{1}(t)$ for position measurement and $r_{2}(t)$ for momentum measurement results in the following stochastic quantum evolution of the quantum harmonic oscillator state (in dimensionless units where $\hat{x}\rightarrow \sqrt{m\omega/\hbar}\hat{x}$, $\hat{p}\rightarrow \hat{p}/\sqrt{\hbar m \omega}$, $t\rightarrow \omega t$)~\cite{karmakar_stochastic_2021}:

\begin{eqnarray}
\frac{dq_{1}}{dt}&=&q_{2}+\frac{q_{3}}{2\tau_{1}}(r_{1}-q_{1})+\frac{q_{4}}{2\tau_{2}}(r_{2}-q_{2}),\nonumber\\
\frac{dq_{2}}{dt}&=&-q_{1}+\frac{q_{4}}{2\tau_{1}}(r_{1}-q_{1})+\frac{q_{5}}{2\tau_{2}}(r_{2}-q_{2}),\nonumber\\
    \frac{dq_{3}}{dt}&=&2q_{4}-\frac{q_{3}^{2}}{2\tau_{1}}-\frac{q_{4}^{2}}{2\tau_{2}}+\frac{1}{2\tau_{2}},\nonumber\\\frac{dq_{4}}{dt}&=&q_{5}-q_{3}-\frac{q_{3}q_{4}}{2\tau_{1}}-\frac{q_{4}q_{5}}{2\tau_{2}},\nonumber\\
      \frac{dq_{5}}{dt}&=&-2q_{4}-\frac{q_{4}^{2}}{2\tau_{1}}-\frac{q_{5}^{2}}{2\tau_{2}}+\frac{1}{2\tau_{1}}.\label{deq}
\end{eqnarray}
The covariances are labeled by, $q_{3}=2(\langle \hat{x}^{2}\rangle-\langle \hat{x}\rangle^{2}),~q_{4}=\langle \hat{x}\hat{p}+\hat{p}\hat{x}\rangle-2\langle \hat{x}\rangle\langle \hat{p}\rangle$, and $q_{5}=2(\langle \hat{p}^{2}\rangle-\langle \hat{p}\rangle^{2})$. The readouts are stochastic variables, $r_{i}=q_{i}+\sqrt{\tau_{i}}\zeta_{i}$, where $\zeta_{i}$ are a Gaussian white noise (due to quantum fluctuations) satisfying $\langle\zeta_{i}(t)\zeta_{i}(0)\rangle=\delta(t)$ and $\tau_{i}$ are the characteristic measurement times, defined as inverses of the corresponding measurement rates~\cite{karmakar_stochastic_2021,chantasri_action_2013,chantasri_stochastic_2015,weber_mapping_2014,korotkov_continuous_1999}.  Work can be extracted in the form of an instantaneous linear feedback Hamiltonian, $H_{\text{fb}}=f_{1}(t)\hat{x}+f_{2}(t)\hat{p}$, where $f_{2}(t)=-\frac{q_{3}}{2\tau_{1}}[r_{1}(t)-\bar{q}_{1}(t)]-\frac{q_{4}}{2\tau_{2}}[r_{2}(t)-\bar{q}_{2}(t)]$ and $f_{1}(t)=\frac{q_{4}}{2\tau_{1}}[r_{1}(t)-\bar{q}_{1}(t)]+\frac{q_{5}}{2\tau_{2}}[r_{2}(t)-\bar{q}_{2}(t)]$(see Appendix E). Here $\bar{q}_{i},~i=1,2$ are the predicted evolution of $q_{i}$ in the absence of measurements, given by, $\bar{q}_{1}(t)=q_{1}(0)\cos{t}+q_{2}(0)\sin{t}$ and $\bar{q}_{2}(t)=-q_{1}(0)\sin{t}+q_{2}(0)\cos{t}.$  A comparable measurements and feedback scheme was also used in Ref.~\cite{flywheel} to stabilize a quantum analogue of flywheel that stores energy  like a battery in its rotational motion.  

We restrict to the case when $\tau_{1}=\tau_{2}=\tau$ which produces coherent states of the quantum oscillator as measurement outcomes and maximizes the engine's efficiency (see Appendix G).   For the covariance matrix initialized in its normal form, $V=\text{diag}\{\nu,\nu\}$ with a corresponding mean number of thermal photons $\bar{n}=\nu-1/2$, the quantum measurement induced evolution is such that it preserves the normal form of the covariance matrix~\cite{karmakar_stochastic_2021}. 
The work along an individual trajectory obeys (in the Stratonovich form),
\begin{eqnarray}
    \frac{d(W/\hbar\omega)}{dt}&=&q_{1}\dot{q}_{1}+q_{2}\dot{q}_{2}=q_{1}\bigg[\frac{q_{3}}{2\sqrt{\tau}}\zeta_{1}(t)+\frac{q_{4}}{2\sqrt{\tau}}\zeta_{2}(t)\bigg]\nonumber\\&+&q_{2}\bigg[\frac{q_{4}}{2\sqrt{\tau}}\zeta_{1}(t)+\frac{q_{5}}{2\sqrt{\tau}}\zeta_{2}(t)\bigg].
\end{eqnarray}
The probability distribution of work extracted at arbitrary finite time is given by,
\begin{equation}
    \mathcal{P}(W/\hbar\omega,t)=\frac{\tau }{\sigma(t)}\exp \left[-\frac{\tau W}{\sigma(t)\hbar\omega}\right],\label{pwork}
\end{equation}
so the work extracted on an average is given by $\langle W/ \hbar\omega\rangle =\sigma(t)/\tau$. The parameter $\sigma(t)$ equals $\nu^{2}(t)dt$ if work is extracted after every measurement of duration $dt$. If the controller decides to apply feedback only after a duration $t$, the work distribution corresponds to the case $\sigma(t)=\int_{0}^{t}dt' \nu^{2}(t')$ in Eq.~\eqref{pwork}. The nonzero average work results from rectifying the quantum noise in the measurement process, and is nonzero even at zero temperature  (see Appendix F). The average power J of the quantum measurement engine is given by,
\begin{equation}
\text{J}(t)=d\langle W/\hbar\omega\rangle/dt=\nu^{2}(t)/\tau,
    \end{equation}
which serves as a useful quantity to infer the relation between rate of information acquisition (the measurement rate) and work extraction for the continuous measurement engine; work is extracted at a faster rate as the measurement rate $\tau^{-1}$ is increased.
 Further, a larger $\nu(t)$ corresponding to a higher thermal quanta in the initial state also allows higher power.

 \subsection{Steady state of the engine}The dynamical equations which describe the evolution of covariance matrix elements are deterministic, and they achieve the steady state value, $\underset{t\gg\tau}{\text{lim}}~\nu(t)=1/2$. In this limit---which is also the steady state of the quantum measurement engine---the quantum measurement dynamics describes coherent state diffusion. The measurement engine also achieves unit efficiency in its steady state; since the measurement merely displaces the ground state (where $\nu(t)=1/2$), the feedback resets the engine perfectly, closing the engine's cycle. 

\subsection{Results}
The characteristics of the continuous quantum measurement engine are shown in Fig.~\ref{Fig2}. We first consider a situation where work is extracted by applying a feedback after continuously measuring the oscillator for a finite duration $t$. Figure~\ref{Fig2}(a) displays a simulation of a single such trajectory, which undergoes two types of dynamics; while the variance decreases to its minimum uncertainty deterministically, the average displacement diffuses as energy is added through the measurement. After a time $t$ the oscillator is reset to extract work. Figure~\ref{Fig2}(b) displays the probability distribution of extracted work from the feedback step. The average work extracted [Fig.~\ref{Fig2}(c)] increases with the duration of the measurement. Subsequently, we consider the situation where work is extracted after each step of the measurement. Figure~\ref{Fig2}(d) displays the average position and momentum of the oscillator after each measurement, which are then reset to the origin by the feedback. Figure~\ref{Fig2}(e) displays the probability distribution of extracted work. The average power J delivered by the quantum engine in unit of the measurement rate $\tau^{-1}$ is shown in Fig.~\ref{Fig2}(f), which decreases and reaches its steady state value as the quantum oscillator is purified by measurements.

\section{Conclusions}We have characterized a quantum engine fueled by simultaneous quantum measurements of both position and momentum observables of a simple harmonic oscillator. We discussed two protocols for operation of the engine, respectively powered by single-shot and continuous quantum measurements. In both cases, the measurement produces a displaced ground state of the quantum oscillator, and work is extracted by shifting the bottom of the harmonic trap suddenly. When compared to their classical counterpart, the quantum engines yield non-zero work output even at zero temperature, 
 demonstrating the energetic consequence of the quantum of noise  added when both quadratures are measured simultaneously; at zero temperature the nonzero work output results from the feedback utilizing the quantum of energy inserted by the phase preserving measurement~\cite{arthurs_simultaneous_1965,loudon_squeezed_1987,clerk_introduction_2010,caves_quantum_1982,caves_quantum_2012,bergeal_phase-preserving_2010,haus_quantum_1962}.  We also derived exact analytical expressions for the probability distributions of extractable work in both transient and steady state of the quantum engine. The availability of exact probability distributions of extractable work may further aid research towards thermodynamically characterizing quantum measurement engines, and derive quantum fluctuation theorems for quantum engines and refrigerators fueled solely by the quantum measurement process~\cite{manikandan_fluctuation_2019,jayaseelan_quantum_2021,elouard_role_2017,buffoni_quantum_2019,elouard_extracting_2017}.

\noindent\textit{Note added.---}Towards the completion of this work, we became aware of a closely related preprint ~\cite{opatrny_work_2021} investigating work extraction from thermal resources using phase sensitive measurements, as opposed to the phase preserving measurements discussed in the present article.

\section{Acknowledgements}This work was supported by the John Templeton Foundation Grant No. 61835. The work of A.A. was supported by the Foundational Questions Institute Fund (Grant No. FQXi-IAF19-05), the Foundational Questions Institute Fund (Grant No. FQXi-IAF19-01) and the ANR Research Collaborative Project ``Qu-DICE" (Grant No. ANR-PRC-CES47). The work of S.K.M. was supported in part by the Wallenberg Initiative on Networks and Quantum Information (WINQ). Nordita is partially supported by Nordforsk. We acknowledge fruitful discussions with Tathagata Karmakar, Philippe Lewalle, Masahito Ueda, Benjamin Huard, R\'{e}my Dassonneville, and L\'{e}a Bresque.
\widetext
\newpage
\appendix
 
 \begin{figure}
\includegraphics[width=0.5\linewidth]{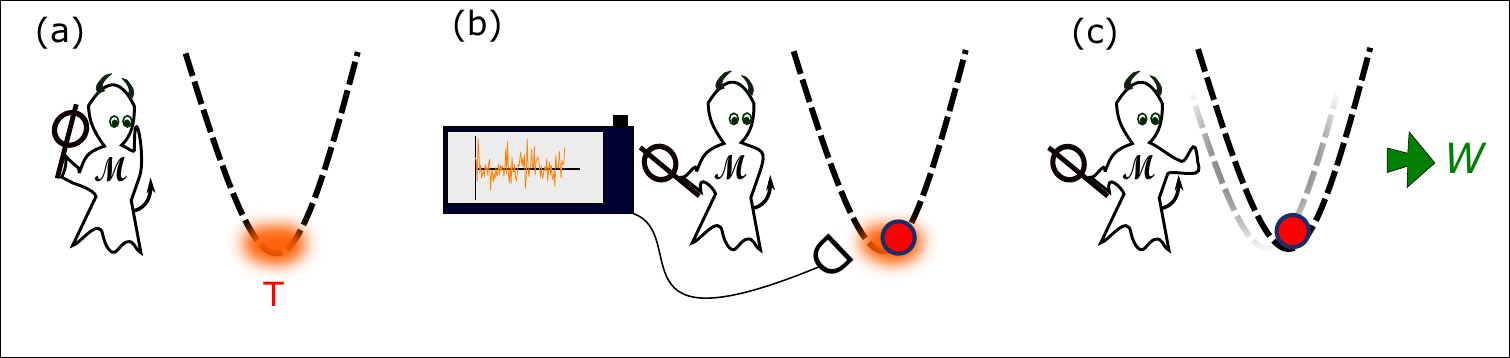}\caption{Cyclic operation of the classical Brownian heat engine~\cite{ashida_general_2014,paneru_lossless_2018}. (a) The Brownian particle in a harmonic trap is initially thermalized to a heat bath at temperature T. (b) The demon measures position of the Brownian particle. (c) Work is extracted by displacing the trap conditioned on the measurement outcome.
\label{Fig3}}
\end{figure}
\section{The Brownian heat engine}
The classical limit of the engine is described by a Brownian particle in a harmonic trap, whose position $x$ is distributed according to the equilibrium distribution $\mathcal{P}(x)=\sqrt{\frac{k}{2\pi k_{B}\text{T}}}\exp\big(-\frac{kx^{2}}{2 k_{B}\text{T}}\big)$, where $k$ is the spring constant, $k_{B}$ is Boltzmann's constant and T is the temperature of the heat bath (see Fig.~\ref{Fig3})~\cite{ashida_general_2014,paneru_lossless_2018}. A Maxwell's demon extracts work from the heat bath by measuring the particle's position $x$, followed by shifting the bottom of the harmonic trap suddenly to the new position of the particle. If the measurement performed is error free with ideal feedback, it is possible to convert all the available information to extractable work,  $W=k x^{2}/2$~\cite{ashida_general_2014}. The demon in this case extracts work on an average, $\langle W\rangle = k_{B}\text{T}/2$ per cycle, which also saturates the achievable upper-bound for extractable work by moving the trap in the classical Brownian engine limit~\cite{ashida_general_2014}. 

\section{Energy contributed by measurement}
The unconditional state of the quantum oscillator after a coherent state basis measurement (implemented by a simultaneous weak measurement of both position and momentum quadratures,  also known as a phase-preserving measurement) can be written as,
\begin{equation}
    \rho_{A}=\int d^{2}\alpha \text{Q}(\alpha)|\alpha\rangle\langle\alpha|,
\end{equation}
where 
$    \text{Q}(\alpha) =\frac{1}{\pi(1+\bar{n})}e^{-|\alpha|^{2}/(1+\bar{n})},$ is the Husimi Q distribution~\cite{husimi_formal_1940} of a thermal state (denoted as $\mathcal{P}_{\text{Q}}$ in Eq.~(1) of the main text) with average number of thermal quanta $\bar{n}$. We can now compute the average number of quanta in the unconditional post-measurement state $\rho$, given by,
\begin{equation}
    \langle \hat{N}\rangle_{A} = \textit{tr}\{\rho_{A}\hat{N}\}=\text{tr}[\rho_{A} a^{\dagger}a]=\int d^{2}\alpha \text{Q}(\alpha)|\alpha|^{2}=\bar{n}+1.
\end{equation}
We find that the measurement, on an average, adds one quantum of energy to the oscillator, which is then perfectly extracted in the feedback step.  The additional quantum in $\text{Q}(\alpha)$ stems from the fact that the coherent state basis is overcomplete; even measurement of the vacuum state will in general yield a coherent state with finite $\alpha$.

Note that the addition of single quantum is a generic property of coherent state basis measurements which can be implemented by a simultaneous weak measurement of both position and momentum of the quantum particle~\cite{arthurs_simultaneous_1965}.  To see this,  we can write an arbitrary initial quantum state $\rho$ in the P representation as using the coherent state basis $|\beta\rangle$,
\begin{equation}
    \rho=\int d^{2}\beta\text{P}(\beta)|\beta\rangle\langle\beta|.
\end{equation}
The P representation can be derived from the initial density matrix $\rho$ in different equivalent ways, for instance see Ref.~\cite{mehta_diagonal_1967}, where it is defined as,
\begin{equation}
    \text{P}(\beta)=\frac{e^{|\beta|^{2}}}{\pi^{2}}\int \langle -\gamma|\rho|\gamma\rangle \exp(|\gamma|^{2}+ \beta\gamma^{*}-\beta^{*}\gamma)d^{2}\gamma,
\end{equation}
where $|\gamma\rangle$ are again coherent states. The different equivalent P representations of $\rho$ are constrained to obey the optical equivalence theorem for the density matrix $\rho$ for the expectation value of any normally ordered operator: $\langle (a^{\dagger})^{n}a^{m}\rangle = \text{tr}[\rho (a^{\dagger})^{n}a^{m}] = \int d^{2}\beta\text{P}(\beta)(\beta^{*})^{n}\beta^{m}$. As an example, the average number of quanta in the initial density matrix $\rho$ maybe computed using the P function as, $\langle \hat{N}\rangle = \langle a^{\dagger}a\rangle = \text{tr}[\rho a^{\dagger}a] = \int d^{2}\beta\text{P}(\beta)|\beta|^{2}=\bar{n}$. The P function for a coherent state $|\alpha\rangle\langle\alpha|$ is the delta function, $\text{P}(\beta)=\delta(\beta-\alpha)$ and a P representation for a thermal state with average number of thermal quanta $\bar{n}$ is,  $\text{P}(\beta)=\frac{1}{\pi\bar{n}}\exp(-|\beta|^{2}/\bar{n})$.

On the other hand, performing a measurement in the coherent state basis produces coherent states $|\alpha\rangle$ with probability $\text{Q}(\alpha)=\frac{1}{\pi}\langle\alpha|\rho|\alpha\rangle$, which is known as the Q representation of the density matrix $\rho$. The unconditional post measurement state $\rho_{A}$ can be written as the following ensemble where the probabilities are given by the Q function~\cite{schleich_quantum_2011,husimi_formal_1940},
\begin{equation}
    \rho_{A}=\int d^{2}\alpha\text{Q}(\alpha)|\alpha\rangle\langle\alpha|.
\end{equation}
So a method to probe the Q function of a quantum state experimentally is to perform measurements in the coherent state basis~\cite{caves_quantum_2012}. We now proceed to compute the average quanta in the post measurement state. In order to do that, we can relate the Q function (which appear as the probability density describing the post measurement state) to the P function via the transformation rule~\cite{schleich_quantum_2011}: $\text{Q}(\alpha)=\frac{1}{\pi}\int d^{2}\beta \text{P}(\beta) e^{-|\alpha-\beta|^{2}}$, and write,
\begin{equation}
    \rho_{A}=\int d^{2}\alpha\bigg(\frac{1}{\pi}\int d^{2}\beta \text{P}(\beta) e^{-|\alpha-\beta|^{2}}\bigg)|\alpha\rangle\langle\alpha|.
\end{equation}
The average number of quanta in the unconditional post measurement state is, $\langle\hat{N}\rangle_{A}=\text{tr}[\rho_{A} a^{\dagger}a].$ This can be evaluated by changing the order of integrals as,
\begin{equation}
    \langle\hat{N}\rangle_{A}=\text{tr}[\rho_{A} a^{\dagger}a]=\frac{1}{\pi}\int d^{2}\beta \text{P}(\beta)\int d^{2}\alpha e^{-|\alpha-\beta|^{2}}|\alpha|^{2}=\int d^{2}\beta \text{P}(\beta)(1+|\beta|^{2})=1+\bar{n},
\end{equation}
where $\bar{n}$ is the average quanta in the state prior to measurement. Clearly, this extra quantum comes from the measurement process, and the addition of a quantum of noise is in the same spirit as how it is typically described as added to the variance of quadratures after the measurement, as discussed in Refs.~\cite{arthurs_simultaneous_1965,caves_quantum_2012}. Here we look at the change in the mean quanta instead, which is the observable relevant to the engine's energetics.  

\section{Binary feedback}
\begin{figure}
\includegraphics[width=0.5\linewidth]{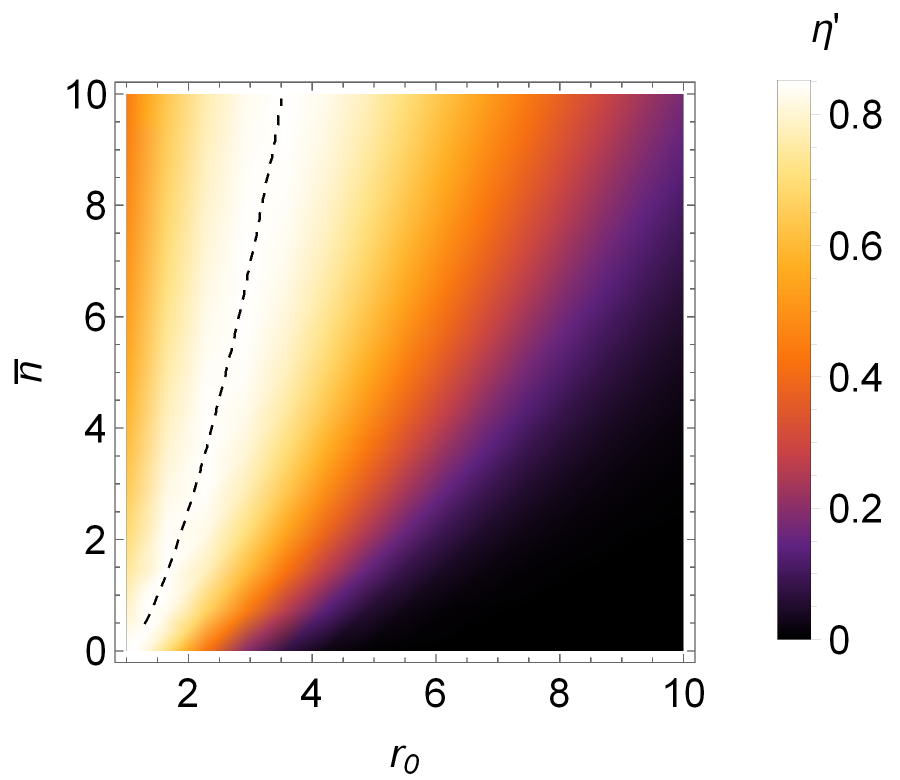}\caption{The efficiency of the engine in the binary feedback $\eta'=\langle W'\rangle/\mathcal{Q}$, as a function of the average number of thermal photons in the initial state $\bar{n}$ and $r_{0}$. The dashed black line indicates the contour line of maximum $\eta': \eta'_{\text{max}}(\bar{n},r_{0})\approx 0.85$.\label{Fig4}}
\end{figure}

The engine's implementation with single shot measurements and a binary-valued feedback is as follows:
\begin{itemize}    \item {Step (a)--(b) are identical to that discussed in the main text.}
\item{Step (c) (work extraction):  When the amplitude of the measured coherent state is greater than $r_{0}/2$, work is extracted by shifting the trap to $r_{0}$, and the system is let to thermalize. Otherwise, no action is performed. The work extracted is zero if $r<r_{0}/2$ and the work extracted is $\hbar\omega(2rr_{0}-r_{0}^{2})$ for $r\geq r_{0}/2$. }
\end{itemize}
The average work extracted in this protocol is,
\begin{eqnarray}
   \langle W' \rangle&=& \hbar\omega\int_{0}^{2\pi} d\theta\int_{r_{0}/2}^{\infty} rdr \mathcal{P}(r,\bar{n})(2rr_{0}-r_{0}^{2})\nonumber\\ &=& \hbar\omega r_{0}\sqrt{\pi(1+\bar{n})}~\text{erfc}\frac{r_{0}}{2\sqrt{1+\bar{n}}}.
\end{eqnarray}
Here $\text{erfc}(u)=2\pi^{-1/2}\int_{u}^{\infty}e^{-y^{2}}dy.$ The efficiency is less than that of the continuous feedback scheme,
  $  \eta'=\frac{\langle W' \rangle}{\mathcal{Q}}<1,$
and has a maximum, $\eta'_{\text{max}} \approx 0.85$ (Fig.~\ref{Fig4}).  
The classical analogue of this protocol is presented in Ref.~\cite{paneru_lossless_2018}.

\section{Cost of memory erasure and the efficiency of the engine}

Here we first discuss Landauer erasure in the Binary feedback protocol discussed above. The memory in this example behaves essentially like a logical bit, with states $L$ or $R$, indicating if the particle is to the left of $r_{0}/2$ or to the right of $r_{0}/2$. The minimum energy cost to erase the memory can be accounted by the Landauer's bound as~\cite{maruyama_colloquium_2009},
\begin{equation}
    W_{\text{res}}' = k_{B}\text{T}_{D}H(p_{0}).
\end{equation}
Here $H(p_{0})=-p_{0}\log{p_{0}}-(1-p_{0})\log{(1-p_{0})}$, where we have defined $p_{0}$ as the concatenated probability of finding the particle to the right of $r_{0}/2$:
\begin{equation}
    p_{0}=\int d\theta\int^{\infty}_{r_{0}/2}rdr\mathcal{P}(r,\bar{n})=e^{-\frac{r_{0}^{2}}{4(\bar{n}+1)}}.
\end{equation}
The Shannon entropy is indeed bounded from above by $\log{2}$ as expected for a classical bit memory. The efficiency $\eta'_{\text{T}}$ of the thermodynamic cycle can now be computed,
\begin{eqnarray}
    \eta'_{\text{T}} &=& \frac{    \langle W'\rangle-W_{\text{res}}'}{\mathcal{Q}} =\frac{    \langle W'\rangle-k_{B}\text{T}_{D}H(p_{0})}{\mathcal{Q}}.
\end{eqnarray}

We now look at continuous measurement examples where at each step both the information stored, and the feedback are continuous-valued. The associated probability density is $\mathcal{P}(r_{1},r_{2})$, where $r_{1},r_{2}$ are the readout variables. As an extension of the above example, we consider a countable discretization for the probability distribution of the readouts $ \bar{\mathcal{P}}(r_{1i},r_{2j})$ where the sampling is done according to bins around points $r_{1i},r_{2j}$ having bin area $\delta^{2}$ such that,
\begin{equation}
    \bar{\mathcal{P}}(r_{1i},r_{2j})\delta^{2}=\int_{i\delta}^{(i+1)\delta}\int_{j\delta}^{(j+1)\delta}dr_{1}dr_{2}\mathcal{P}(r_{1},r_{2}),
\end{equation}
by the mean value theorem~\cite{cover_elements_2012}. Here $\delta^{2}$ can be related to the resolution of the detector. The entropy of this discrete probability distribution is given by,
\begin{eqnarray}
   H(\bar{\mathcal{P}})=-\sum_{i,j} \delta^{2}\bar{\mathcal{P}}(r_{1i},r_{2j})\log{[ \bar{\mathcal{P}}(r_{1i},r_{2j})\delta^{2}]}=-\sum_{i,j} \delta^{2}\bar{\mathcal{P}}(r_{1i},r_{2j})\log{[ \bar{\mathcal{P}}(r_{1i},r_{2j})]}-\sum_{i,j} \delta^{2}\bar{\mathcal{P}}(r_{1i},r_{2j})\log{\delta^{2}}.
\end{eqnarray}
Assuming $\mathcal{P}(r_{1},r_{2})\log{ \mathcal{P}(r_{1},r_{2})}$ is Riemann integrable, we note that the first term $-\sum_{i,j} \delta^{2}\bar{\mathcal{P}}(r_{1i},r_{2j})\log{ \bar{\mathcal{P}}(r_{1i},r_{2j})}\rightarrow -\int dr_{1}dr_{2}\mathcal{P}(r_{1},r_{2})\log{ \mathcal{P}(r_{1},r_{2})}=H(\mathcal{P})$ when $\delta\rightarrow 0$. We also note that $\sum_{i,j} \delta^{2}\bar{\mathcal{P}}(r_{1i},r_{2j})=1$. Therefore for a discretization into squares of area $\delta^{2}$, the associated Shannon entropy scales approximately as, $H(\bar{\mathcal{P}})=H(\mathcal{P})-\log{\delta^{2}}$~\cite{cover_elements_2012}. The efficiency of the thermodynamic cycle at each step becomes, 
\begin{eqnarray}
    \eta_{\text{T}} &=& \frac{    \langle W\rangle-W_{\text{res}}}{\mathcal{Q}} =\frac{    \langle W\rangle-k_{B}\text{T}_{D}H(\bar{\mathcal{P}})}{\mathcal{Q}},
\end{eqnarray}
 where $\langle W\rangle$ is the average work extracted at each step. 
\section{Continuous feedback}
The work extraction protocol is essentially a linear feedback stabilization protocol for the quantum oscillator. The effect of a linear Hamiltonian  $H_{\text{fb}}=f_{1}\hat{x}+f_{2}\hat{p}$ on the quantum mechanical averages of $\hat{x}$ and $\hat{p}$ when measurements are done continuously is,
\begin{eqnarray}
\frac{dq_{1}}{dt}&=&q_{2}+\frac{q_{3}}{2\tau_{1}}(r_{1}-q_{1})+\frac{q_{4}}{2\tau_{2}}(r_{2}-q_{2})+f_{2},~~
\frac{dq_{2}}{dt}=-q_{1}+\frac{q_{4}}{2\tau_{1}}(r_{1}-q_{1})+\frac{q_{5}}{2\tau_{2}}(r_{2}-q_{2})-f_{1},
   \label{deqfb}
\end{eqnarray}
where $q_{1}=\langle \hat{x}\rangle$ and $q_{2}=\langle \hat{p}\rangle$. Therefore we can choose the amplitude of the linear feedback Hamiltonian, $f_{2}(t)=-\frac{q_{3}}{2\tau_{1}}[r_{1}(t)-\bar{q}_{1}(t)]-\frac{q_{4}}{2\tau_{2}}[r_{2}(t)-\bar{q}_{2}(t)]$ and $f_{1}(t)=\frac{q_{4}}{2\tau_{1}}[r_{1}(t)-\bar{q}_{1}(t)]+\frac{q_{5}}{2\tau_{2}}[r_{2}(t)-\bar{q}_{2}(t)]$ such that the unitary evolution resulting from feedback effectively cancels the measurement back-action on the quantum state. Here $\bar{q}_{i},~i=1,2$ is the predicted evolution of $q_{i}$ in the absence of measurements, given by, $\bar{q}_{1}(t)=q_{1}(0)\cos{t}+q_{2}(0)\sin{t}$ and $\bar{q}_{2}(t)=-q_{1}(0)\sin{t}+q_{2}(0)\cos{t}$, in units where $t\rightarrow \omega t$.

 \section{Work extracted on average in the continuous measurement protocol}
As discussed in the main text, the work extracted along an individual quantum trajectory in the continuous measurement protocol satisfies the following stochastic equation,
\begin{eqnarray}
    \frac{d(W/\hbar\omega)}{dt}&=&q_{1}\dot{q}_{1}+q_{2}\dot{q}_{2}=q_{1}\bigg[\frac{q_{3}}{2\sqrt{\tau}}\zeta_{1}(t)+\frac{q_{4}}{2\sqrt{\tau}}\zeta_{2}(t)\bigg]+q_{2}\bigg[\frac{q_{4}}{2\sqrt{\tau}}\zeta_{1}(t)+\frac{q_{5}}{2\sqrt{\tau}}\zeta_{2}(t)\bigg].
\end{eqnarray}
We further assume initial conditions $q_{1}(0)=q_{2}(0)=0$ and $q_{3}(0)=q_{5}(0),~q_{4}(0)=0$. As a result, the extractable work at arbitrary time $t$ is the integral, $W(t)/\hbar\omega=\int_{0}^{t}dt'\bigg[q_{1}(t')\frac{q_{3}(t')}{2\sqrt{\tau}}\zeta_{1}(t')+q_{2}(t')\frac{q_{5}(t')}{2\sqrt{\tau}}\zeta_{2}(t')\bigg].$ The above integral is cast in the Stratonovich form and can be discretized as,
\begin{equation}
    \int_{0}^{t}dt'\bigg[q_{1}(t')\frac{q_{3}(t')}{2\sqrt{\tau}}\zeta_{1}(t')+q_{2}(t')\frac{q_{5}(t')}{2\sqrt{\tau}}\zeta_{2}(t')\bigg]=\frac{dt'}{2\sqrt{\tau}}\sum_{i=1}^{N-1}\frac{q_{1}^{i}+q_{1}^{i+1}}{2}q_{3}^{i}\zeta_{1}^{i}+\frac{dt'}{2\sqrt{\tau}}\sum_{i=1}^{N-1}\frac{q_{2}^{i}+q_{2}^{i+1}}{2}q_{5}^{i}\zeta_{2}^{i}.
\end{equation}
Here $q_{3}(t'),~q_{5}(t')$ are the quadrature variances which evolve deterministically according to Eq.~(3) of the main text. We can now use the relations, $q_{1}^{i+1}=q_{1}^{i}+dt'\bigg(q_{2}^{i}+\frac{q_{3}^{i}}{2\sqrt{\tau}}\zeta_{1}^{i}\bigg),$ and $q_{2}^{i+1}=q_{2}^{i}+dt'\bigg(-q_{1}^{i}+\frac{q_{5}^{i}}{2\sqrt{\tau}}\zeta_{2}^{i}\bigg)$ (together with the observation that the stochastic averages of terms which are linear in $\zeta$ vanish, and $\zeta_{i}^{2}dt'=1$ by Ito's rule) in order to compute the stochastic average of extractable work. We have,
\begin{eqnarray}
\langle W/\hbar\omega\rangle &=& \bigg\langle   \frac{dt'}{2\sqrt{\tau}}\sum_{i=1}^{N-1}\frac{2q_{1}^{i}+dt'\bigg(q_{2}^{i}+\frac{q_{3}^{i}}{2\sqrt{\tau}}\zeta_{1}^{i}\bigg)}{2}q_{3}^{i}\zeta_{1}^{i}+\frac{dt'}{2\sqrt{\tau}}\sum_{i=1}^{N-1}\frac{2q_{2}^{i}+dt'\bigg(-q_{1}^{i}+\frac{q_{5}^{i}}{2\sqrt{\tau}}\zeta_{2}^{i}\bigg)}{2}q_{5}^{i}\zeta_{2}^{i}\bigg\rangle,\nonumber\\&=&
\frac{dt'}{8\tau}\sum_{i=1}^{N-1}(q_{3}^{i})^{2}+(q_{5}^{i})^{2}\approx \frac{1}{\tau}\int_{0}^{t}dt'\nu(t')^{2}.
\end{eqnarray}
Above we have defined $\nu(t') = q_{3}(t')/2=q_{5}(t')/2$, when $q_{3}(0)=q_{5}(0),~\text{and}~q_{4}(0)=0$. It follows that work is extracted at a rate,~ $d\langle W/\hbar\omega\rangle/dt'=\nu^{2}(t')/\tau.$ The above calculation also shows that the extracted work originates from the noise due to quantum fluctuations in the simultaneous quantum measurement process, with each quadrature measurement channel contributing work at a rate $\nu^{2}(t')/(2\tau)$. 
\subsection*{Ito interpretation}
The extracted work is given by, $W/\hbar\omega = (q_{1}^{2}+q_{2}^{2})/2=f(q_{1},q_{2})$. Using Ito's lemma, we have,
\begin{equation}
    df=\frac{\partial f}{\partial t'}dt'+\frac{\partial f}{\partial q_{1}}dq_{1}+\frac{\partial f}{\partial q_{2}}dq_{2}+\frac{1}{2}\frac{\partial^{2} f}{\partial q_{1}^{2}}dq_{1}^{2}+\frac{1}{2}\frac{\partial^{2} f}{\partial q_{2}^{2}}dq_{2}^{2}+\frac{\partial^{2} f}{\partial q_{1}q_{2}}dq_{1}dq_{2}+ ...\label{ito1}
\end{equation}
Also note that $\frac{\partial f}{\partial t'}=0$,~ $\frac{\partial f}{\partial q_{1}}=q_{1}$,~ $\frac{\partial f}{\partial q_{2}}=q_{2}$,~$\frac{\partial^{2} f}{\partial q_{1}^{2}}=\frac{\partial^{2} f}{\partial q_{2}^{2}}=1$ and $\frac{\partial^{2} f}{\partial q_{1}q_{2}}=0$. We can also use the coordinate equations in the Ito form, given by:
\begin{equation}
    dq_{1}=q_{2}dt'+\frac{q_{3}}{2\sqrt{\tau}}d\mathcal{W}_{1}+\frac{q_{4}}{2\sqrt{\tau}}d\mathcal{W}_{2},~\text{and}~
dq_{2}=-q_{1}dt'+\frac{q_{4}}{2\sqrt{\tau}}d\mathcal{W}_{1}+\frac{q_{5}}{2\sqrt{\tau}}d\mathcal{W}_{2}.
\end{equation}
Here $\mathcal{W}_{i},~~i=1,2.,$ are the Wiener increments. Substituting in Eq.~\eqref{ito1}, we get the following differential,
\begin{eqnarray}
   df&=&q_{1}\bigg(q_{2}dt'+\frac{q_{3}}{2\sqrt{\tau}}d\mathcal{W}_{1}+\frac{q_{4}}{2\sqrt{\tau}}d\mathcal{W}_{2}\bigg)+q_{2}\bigg(-q_{1}dt'+\frac{q_{4}}{2\sqrt{\tau}}d\mathcal{W}_{1}+\frac{q_{5}}{2\sqrt{\tau}}d\mathcal{W}_{2}\bigg)\nonumber\\&+&\frac{1}{2}\bigg[\bigg(q_{2}dt'+\frac{q_{3}}{2\sqrt{\tau}}d\mathcal{W}_{1}+\frac{q_{4}}{2\sqrt{\tau}}d\mathcal{W}_{2}\bigg)^{2}+\bigg(-q_{1}dt'+\frac{q_{4}}{2\sqrt{\tau}}d\mathcal{W}_{1}+\frac{q_{5}}{2\sqrt{\tau}}d\mathcal{W}_{2}\bigg)^{2}\bigg]+ ...
\end{eqnarray}
We are interested in the case when $q_{3}(0)=q_{5}(0),~\text{and}~q_{4}(0)=0$ [leading to $q_{3}(t')=q_{5}(t')=2\nu(t'),~\text{and}~q_{4}(t')=0$]. Further we use $d\mathcal{W}_{i}^{2}=dt'$ and keep terms of order $dt'$ or less to obtain,
\begin{eqnarray}
   df=q_{1}\frac{q_{3}}{2\sqrt{\tau}}d\mathcal{W}_{1}+q_{2}\frac{q_{5}}{2\sqrt{\tau}}d\mathcal{W}_{2}+\frac{1}{2}\bigg(\frac{q_{3}^{2}}{4\tau}dt'+\frac{q_{5}^{2}}{4\tau}dt'\bigg)=\frac{\nu(t')^{2}}{\tau}dt'+\frac{q_{1}q_{3}}{2\sqrt{\tau}}d\mathcal{W}_{1}+\frac{q_{2}q_{5}}{2\sqrt{\tau}}d\mathcal{W}_{2}.
\end{eqnarray}
The drift terms vanish upon stochastic average, demonstrating that the average power J delivered by the engine is $\text{J}(t')=\frac{\nu(t')^{2}}{\tau}$. the unit of energy, $\hbar \omega,$ is implicit in the energy (power) definition, J.

\section{Efficiency of the engine with continuous feedback}
\begin{figure}
\includegraphics[width=0.5\linewidth]{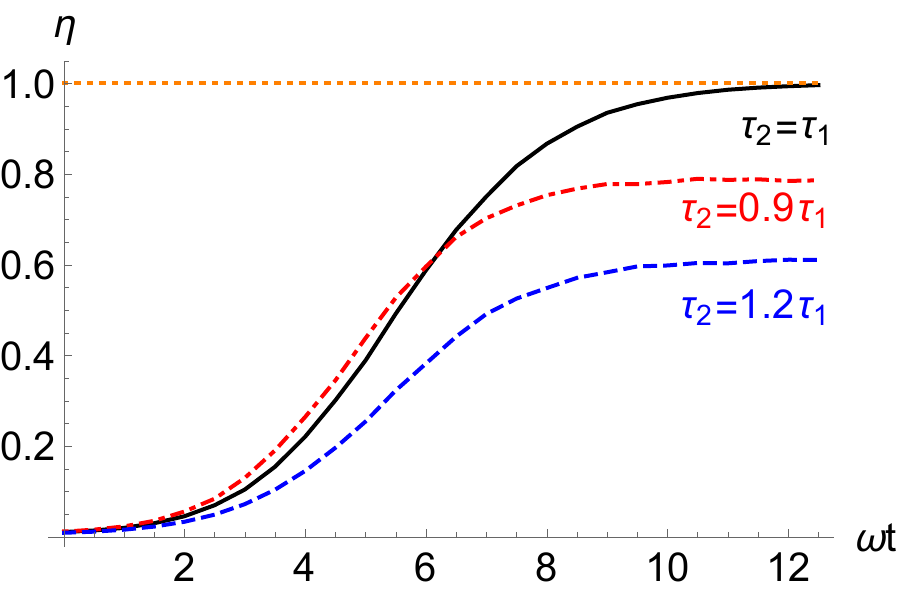}\caption{The efficiency of the engine in three cases, $\tau_{2}=\tau_{1}$, $\tau_{2}=0.9\tau_{1}$, and $\tau_{2}=1.2\tau_{1}$, averaged over $10^4$ quantum trajectories. We note that unit efficiency is reached in the steady state when $\tau_{2}=\tau_{1}$.  \label{Fig5}}
\end{figure}

We define the stochastic energy of the quantum oscillator as $\mathcal{Q}/\hbar\omega=H/\hbar\omega-1/2=(\langle p^{2}\rangle+\langle x^{2}\rangle)/2 =(\text{var}(x)+\text{var}(p)+\langle x\rangle^{2}+\langle p\rangle^{2})/2-1/2=(q_{1}^{2}+q_{2}^{2})/2+(q_{3}+q_{5})/4-1/2=W/\hbar\omega+(q_{3}+q_{5})/4-1/2.$ We have identified the energy stored in the displacement as extractable work, $W$. The subtracted half corresponds to the zero-point energy of the quantum oscillator. When $\tau_{1}=\tau_{2}=\tau$, $q_{3}(t)=q_{5}(t)=2\nu(t),~\text{and}~q_{4}(t)=0$. In this case we obtain, 
\begin{equation}
   \mathcal{Q}/\hbar\omega=W/\hbar\omega +4\nu(t)/4-1/2 \xrightarrow{t\gg\tau} W/\hbar\omega,
\end{equation}
since in the steady state ($t\gg\tau$) we have $\underset{t\gg\tau}{\text{lim}}~\nu(t)=1/2$. We thus notice that in the continuous feedback case, when work is extracted after each measurement, the extracted work $W\rightarrow \mathcal{Q}$ in the long time limit, demonstrating that the work conversion efficiency of the engine $\eta=W/\mathcal{Q}\rightarrow 1$. Here the energy extracted as work differs from the average Hamiltonian change for the quantum oscillator in the transient regime, which includes energy stored in the variances of $\hat{x}$ and $\hat{p}$ that evolve according to Eq.~(3). Moving the bottom of the trap---which extracts part of the average Hamiltonian change as work---preserves the variances. Extracted work becomes equal to the average Hamiltonian change in the steady state, when the variances reach their fixed point values. The efficiency of the engine in three cases, $\tau_{2}=\tau_{1}$, $\tau_{2}=0.9\tau_{1}$, and $\tau_{2}=1.2\tau_{1}$ are shown in Fig.~\ref{Fig5}. We note that unit efficiency is reached in the steady state when $\tau_{2}=\tau_{1}$.

  \bibliography{SHOengine.bib}
		   \end{document}